# Correlation radius in thin ferroelectric films

M.D. Glinchuk, A.N. Morozovska [*] and E.A. Eliseev

Institute for Problems of Materials Science, NAS of Ukraine,

Krjijanovskogo 3, 03142 Kiev, Ukraine, glin@materials.kiev.ua

**Abstract**

In the paper we present analytical calculations of the profiles and average values correlation radius of polarization fluctuations and generalized susceptibility in thin ferroelectric films of thickness $L$ with in-plane (a-films) and out-of-plane (c-films) polarization orientation. The contribution of polarization gradient, surface energy and depolarization field (if any) were taken into account. For the second order ferroelectrics the correlation radius and generalized susceptibility diverge at the film critical thickness $L_{cr}$ as anticipated. In the case of a-films where depolarization field is absent, the surface energy and polarization gradient govern the size effects such as thickness-induced phase transition, corresponding correlation radius and generalize susceptibility divergence in the phase transition point. In the case of c-films with strong depolarization field, the surface energy, polarization gradient, intrinsic depolarization effect related with polarization inhomogeneous distribution and incomplete screening in the electrodes determine the correlation radius and generalized susceptibility size effect. At that contribution of all the factors are comparable at reasonable material parameters, but could be easily tuned by varying of surface energy coefficients, polarization gradient contribution and screening conditions.

## 1. Introduction

Since Landau-Ginzburg-Devonshire (**LGD**) theory is widely used for description of ferroelectric nanosystems [1, 2, 3], its applicability demands estimation of the system sizes range, for which the phenomenological approach is valid. It is known that phenomenological theory describe pretty good long-range order phase in ferroelectrics that can exist either in a whole bulk sample with homogeneous polarization or in the regions of correlation radius $R_c$ size if polarization is inhomogeneous [4]. In nanoferroelectrics the polarization is inhomogeneous by definition because of surface influence, being strongly inhomogeneous in the vicinity of surface ("shell" region) and close to homogeneous polarization of the bulk sample in some central part

---

[*] permanent address: V. Lashkarev Institute of Semiconductor Physics, NAS of Ukraine, 41, pr. Nauki, 03028 Kiev, Ukraine

("core") of thin film or nanoparticle. The ratio of the shell to core volumes increases with the system sizes decrease [5]. For the validity of phenomenological approach to the whole film or nanoparticle one has to chose the strictest condition for ferroelectricity existence with or without correlation effect. Namely it is necessary to have nanosystems with the sizes larger than the critical ones. If not so the ferroelectric long range order might not exist and so phenomenological approach will be invalid.

It was shown earlier, that critical thickness essentially increases with depolarization field increase [6]. This field makes polarization more homogeneous [7] and so the applicability of phenomenological approach at $L \geq L_{cr}$ is out of doubts. The existence of conducting electrodes as well as the domain structure must decrease the depolarization field value and so increase the ferroelectric phase stability and thus $L_{cr}$ has to decrease. So the criteria of phenomenological theory approach validity $L \geq L_{cr}$ will be satisfied in broad enough region of film thickness. It is obvious that correlation radius we considered is the characteristic scale of spontaneous polarization fluctuations and so the existence of polarization is necessary, i.e. $L \geq L_{cr}$ and of course $L_{cr} > R_c$, so we have the inequality $L \geq L_{cr} > R_c$. One can expect that this condition can be valid for ferroelectric films with polarization orientation in the film plane (***a-films***), while for films with polarization orientation normal to the film plane (***c-films***) the region of phenomenological theory validity might be more narrow. In what follows we will considered in details a- and c-films to show that generally the region of phenomenological theory applicability is broad enough, namely starting from film thickness from several to several tens nm.

The applicability of LGD theory is corroborated by the fact, that the critical sizes of long-range order existence in ferroics calculated from microscopic [8] and phenomenological [9] theories are in a good agreement with each other as well as with experimental results for ferromagnetic [10] and ferroelectric [11, 12, 13] systems. Polarization distribution in thin ferroelectric films were studied experimentally [14, 15] and calculated from the first principles [16, 17]. Performed analyses of all these results show that polarization distribution over the film depth is essentially inhomogeneous, at that the inhomogeneity increases with the film thickness decrease. Polarization strongly differs from its average value as well as from the bulk value in ultrathin films («parabolic distribution»).

In worth to mention the papers of Majdoub et al. [18, 19], which consider the influence of flexoelectric effect and polarization gradient on piezoelectric and dielectric properties of different nanosystems, such as cantilevers and thin films. The systems were considered both within the phenomenological theory framework as well as within microscopic theory (ab initio calculations and molecular dynamics simulations). It was proved that the microscopic results in majority of cases even quantitatively and in other cases qualitatively are reproduced by the phenomenological calculations, at that polarization gradient plays the leading role.

In contrast to Majdoub et al papers, which correctly apply the LGD phenomenology to nanosystems, Tagantsev et al. [20] seem to pretend on «making the bridges» between phenomenological theory and microscopic calculations from the first principles. However Tagantsev et al groundlessly assumed that the surface energy is determined by the polarization value in the bulk of the sample and did not consider at all the polarization gradient contribution into the system free energy. Actually such assumption not only makes impossible to consider the spatial distribution of the physical quantities (in particular polarization and depolarization field) but also provides unjustified results about their averaged values. The effects of ferroelectric film splitting on 180-degree stripe domains, which appear under the condition of incomplete screening in the electrodes (e.g. semiconducting electrodes with finite screening length or in the presence of dielectric dead layer), is well-known [21, 22] and typically energetically preferable. However, Tagantsev et al even do not mention the effect of domain splitting. It is well-known that transverse polarization gradient cannot be neglected anyway, since under the ignorance of the positive gradient energy the appearance of stripe domains is always energetically preferable at that the stripe period tends to zero in order to minimize depolarization field energy. Moreover, Tagantsev et al estimate the shell region thickness near the $BaTiO_3$ film interfaces as 0.2 nm basing on the neutron scattering data obtained for bulk samples. However for nanosized ferroelectrics the shell thickness appeared much higher, e.g. precise measurements [23, 24] of the domain wall width (5 ÷ 2.5 nm) in $LiTaO_3$ films of thickness ~ 50 ÷ 500 nm, give estimation for the shell as ~5 nm. The value can be even much higher in soft ferroelectrics like Roshelle salt, $BiFeO_3$ or several $Pb(Zr,TiO_3)$ compositions. Thus unjustified strong inequality "shell scale << lattice constant" used by Tagantsev et al in order to neglect the polarization gradient and do not consider the polarization distribution is invalid for the majority of ferroelectric nanosystems.

Really, the polarization inhomogeneity in ferroelectric nanosystems was proved experimentally in many papers [14, 15, 25] and so existence of polarization gradient and strong surface influence obtained experimental confirmation. Thus, it is obvious that the measurements and adequate calculations of correlation radii and critical thickness seem to be important. On the other hand the knowledge of correlation radius in particles is extremely important for observation of recently forecasted superparaelectric phase in the ensemble of noninteracting ferroelectric nanoparticles [26].

In the paper we present analytical calculations of the correlation radius of polarization fluctuations and generalized susceptibility in thin ferroelectric a- and c-films. For the second order ferroelectrics the correlation radius and generalized susceptibility diverge at the film critical thickness as anticipated. However, ferroelectric phase we are interested in corresponds to the thickness larger than the critical one. In the case of a-films without depolarization field the surface energy and polarization gradient govern the size effects such as thickness-induced phase transition, corresponding correlation radius divergence in the phase transition point and other properties behavior. In the case of c-films with strong depolarization field the surface energy, polarization gradient and intrinsic depolarization effect related with polarization inhomogeneous distribution and incomplete screening of depolarization field even in superconducting electrodes determine the correlation radius and generalized susceptibility size effect. At that contribution of all the factors are comparable at reasonable material parameters, but could be easily tuned by varying of surface energy coefficients, polarization gradient coefficient and screening conditions.

Obtained results prove that oversimplifications of the phenomenological approach (such as neglecting the polarization gradient and/or unjustified assumption about negligibly small surface energy contribution) lead to invalid conclusions about the contributions of different physical mechanisms into the size effect of thin ferroelectric films and to the wrong estimations of phenomenological theory applicability to nanosystems.

**2. Basic definitions**

Let us introduce a correlation function of one-component spontaneous polarization $P(\mathbf{r})$ fluctuations in conventional way [27, 28]

$$G(\mathbf{r},\mathbf{r}') = \langle (P(\mathbf{r}) - \langle P(\mathbf{r}) \rangle)(P(\mathbf{r}') - \langle P(\mathbf{r}') \rangle) \rangle, \quad (1)$$

where $\langle ... \rangle$ stands for thermal (statistical) averaging. Using the fluctuation-dissipation theorem, one can rewrite the correlation function (1) via a generalized susceptibility $\chi(\mathbf{r},\mathbf{r}')$ in the form $G(\mathbf{r},\mathbf{r}') = k_B T \chi(\mathbf{r},\mathbf{r}')$ [28], where $\chi(\mathbf{r},\mathbf{r}')$ determines the increment of polarization $\delta P(\mathbf{r})$ under the inhomogeneous electric field $\delta E(\mathbf{r}')$:

$$\delta P(\mathbf{r}) = \int \chi(\mathbf{r},\mathbf{r}') \delta E(\mathbf{r}') d\mathbf{r}'. \quad (2)$$

Hereinafter symbol δ designates small fluctuations.

In order to find the generalized susceptibility $\chi(\mathbf{r},\mathbf{r}') \sim G(\mathbf{r},\mathbf{r}')$ of confined system, one has to consider the equation of state for one-component of spontaneous polarization $P(\mathbf{r}) = P_S(z) + \delta P_S(\mathbf{r})$:

$$\alpha P + \beta P^3 - g\left(\frac{\partial^2 P}{\partial z^2} + \frac{\partial^2 P}{\partial x^2} + \frac{\partial^2 P}{\partial y^2}\right) = E_0 + E_d(P). \quad (3)$$

Gradient term coefficient $g > 0$, expansion coefficient $\beta>0$ for the second order phase transitions considered hereinafter. Coefficient $\alpha(T) = \alpha_T(T - T_c)$, $T_c$ is the transition temperature of bulk material. The coefficient $\beta$ does not critically depend on temperature $T$. $E_0$ is the external electric field. Linear operator $E_d(P)$ represents depolarization field (if any), that exists due to the polarization inhomogeneity in confined system even under the short-circuited conditions. It essentially depends on the system shape and boundary conditions, provided that $E_d(0) \equiv 0$. For the most of the cases $E_d(P)$ has only integral representations, which reduces to constant (depolarization factors) only for special case of ellipsoidal bodies with homogeneous polarization distribution.

The equilibrium polarization $P_S(z)$ of the monodomain film satisfies the nonlinear Euler-Lagrange equation with the boundary conditions:

$$\begin{cases} \alpha P_S + \beta P_S^3 + g\dfrac{d^2}{dz^2}P_S = E_0 + E_S^d(z), \\ \left(P_S - \lambda_1 \dfrac{dP_S}{dz}\right)\bigg|_{z=0} = 0, \quad \left(P_S + \lambda_2 \dfrac{dP_S}{dz}\right)\bigg|_{z=L} = 0. \end{cases} \quad (4)$$

Here subscript s=3,1, $\lambda_{1,2}$ are extrapolation lengths.

For **a-films** with in-plane spontaneous polarization $P_1(z)$ depolarization field $E^d$ is absent, the fact essentially simplify mathematical treatment of Eq.(4). For **c-films** with out-of-plane spontaneous polarization $P_3(z)$, Maxwell's equations $\text{div}\,\mathbf{D}(\mathbf{r}) = 0$ and $\text{rot}\,\mathbf{E}(\mathbf{r}) = 0$ rewritten for electrostatic potential $\varphi$ lead $\varepsilon_{33}^b \dfrac{\partial^2 \varphi}{\partial z^2} + \varepsilon_{11}\left(\dfrac{\partial^2 \varphi}{\partial x^2} + \dfrac{\partial^2 \varphi}{\partial y^2}\right) = \dfrac{1}{\varepsilon_0}\dfrac{\partial P_3}{\partial z}$. Here we introduced dielectric permittivity or reference state [29] as $\varepsilon_{33}^b$ (typically $\varepsilon_{33}^b \leq 10$); transverse dielectric permittivity $\varepsilon_{11} = \varepsilon_{22}$ and universal dielectric constant $\varepsilon_0$. Corresponding depolarization field acquires the form $E_3^d(z) = \dfrac{(1-\eta)\overline{P_3} - P_3(z)}{\varepsilon_0 \varepsilon_{33}^b}$, where the factor $0 \leq \eta \leq 1$ reflects degree of screening in the interface, conducting or semiconducting electrodes [30, 31]. Note that some metallic electrodes and especially superconducting ones could decrease $\eta$ value up to zero and so decrease the depolarization field [30-31]. For the case of dead layers with thickness $H_{1,2}$ with dielectric permittivity $\varepsilon_{g1,2}$ we derived that $\eta = \dfrac{H_1/\varepsilon_{g1} + H_2/\varepsilon_{g2}}{H_1/\varepsilon_{g1} + H_2/\varepsilon_{g2} + L/\varepsilon_{33}^b}$. Minimal depolarization field related with intrinsic polarization gradient appeared even for the superconducting electrodes (i.e. at $\eta = 0$).

Using equation of state (3), one can write the linearized equation for the fluctuation $\delta P$ as:

$$\begin{cases} \left(\alpha + 3\beta P_S^2(z)\right)\delta P_S - g\left(\dfrac{\partial^2}{\partial z^2} + \dfrac{\partial^2}{\partial x^2} + \dfrac{\partial^2}{\partial y^2}\right)\delta P_S = \delta E_S(x,y,z) + E_S^d(\delta P_S) \\ \left.\left(\delta P_S - \lambda_1 \dfrac{d}{dz}\delta P_S\right)\right|_{z=0} = 0, \quad \left.\left(\delta P_S + \lambda_2 \dfrac{d}{dz}\delta P_S\right)\right|_{z=L} = 0. \end{cases} \quad (5)$$

It is clear from the boundary conditions in Eqs.(5) that purely transverse fluctuations $\delta P_S(x,y)$ cannot exist in the film at finite extrapolation lengths, instead the solution should be z-dependent as $\delta P_S(x,y,z)$, while we mainly consider purely transverse perturbation $\delta E_S(x,y)$.

### 3. Correlation radius and generalized susceptibility distribution in a-films

Exact solution for polarization distribution $P_{1,2}(z)$ is listed in Refs. [32, 33, 34]. General approach for the approximate solution derivation was proposed in Refs.[35, 36] within the framework of direct variational method.

Using Fourier transformation over $\{x,y\}$ coordinates, approximate analytical expression for linear susceptibility was derived under the assumption $P_1^2(z) \to \overline{P_1}^2$, where $\overline{P_1}$ is the spontaneous polarization averaged over the film thickness. We obtained that

$$\delta\widetilde{P}_1(\mathbf{k},z) = \int_0^z dz'\,\delta\widetilde{E}_1(\mathbf{k},z')G(R_a(k),z,z') + \int_z^L dz'\,\delta\widetilde{E}_1(\mathbf{k},z')G(R_a(k),z',z). \quad (6)$$

Where the transverse wave vector $\mathbf{k} = \{k_1, k_2\}$ and its absolute value is $k = \sqrt{k_1^2 + k_2^2}$. Radius

$$R_a(k) = \sqrt{\dfrac{g}{\alpha(T) + 3\beta\overline{P_1}^2 + gk^2}} \quad (7)$$

Green function

$$G(R,z,z') = \dfrac{R}{g} \cdot \dfrac{(\lambda_1 \cosh(z/R) + R\sinh(z/R))(\lambda_2 \cosh((L-z')/R) + R\sinh((L-z')/R))}{R(\lambda_1 + \lambda_2)\cosh(L/R) + (R^2 + \lambda_1\lambda_2)\sinh(L/R)} \quad (8)$$

Generalized susceptibility defined from Eqs.(6) as variational derivative [37] leads to the expression

$$\widetilde{\chi}_{11}(k,z) = \dfrac{\partial\delta\widetilde{P}_1(\mathbf{k},z)}{\partial\delta\widetilde{E}_1(\mathbf{k})} = \dfrac{R_a^2(k)}{g} f(z,L,R_a(k)). \quad (9)$$

The spatial distribution (9) is governed by:

$$f(z,L,R) = 1 - R\dfrac{\lambda_1 \cosh(z/R) + R\sinh(z/R) + \lambda_2 \cosh((L-z)/R) + R\sinh((L-z)/R)}{R(\lambda_1 + \lambda_2)\cosh(L/R) + (R^2 + \lambda_1\lambda_2)\sinh(L/R)} \quad (10)$$

The expression for the critical thickness $L_{cr}(T)$ or critical temperature $T_{cr}(L)$ of size-induced paraelectric phase transition can be obtained from the divergence of $\tilde{\chi}_{11}(0,z)$ at $\overline{P}_1^2 \to 0$, namely:

$$L_{cr}(T) = \sqrt{\frac{g}{\alpha(T)}} \operatorname{arctanh}\left(-\frac{\sqrt{g/\alpha(T)}(\lambda_1 + \lambda_2)}{\lambda_1 \lambda_2 + g/\alpha(T)}\right), \tag{11}$$

$$T_{cr}(L) \approx T_C - \frac{\pi^2(g/\alpha_T)}{\pi^2 \lambda_1 L + 2L^2} - \frac{\pi^2(g/\alpha_T)}{\pi^2 \lambda_2 L + 2L^2}. \tag{12}$$

Using direct variational method with the trial function based on the solution of linearized problem (4) for substitution $P_1(z) = \overline{P}_1 + p(z)$, we obtained equilibrium polarization profile self-consistently from the equation $P_1(z) = \left(\dfrac{E_0 - (\alpha \overline{P}_1 + \beta \overline{P}_1^3)}{\alpha + 3\beta \overline{P}_1^2} + \overline{P}_1\right) f(z, L, R_a(k=0))$. The approximate solution for the equilibrium spontaneous polarization averaged over the film thickness is

$$\frac{1}{L}\int_0^L P_1(z)dz \equiv \overline{P}_1 = \sqrt{\frac{-\alpha(T)}{\beta(3-2\overline{f})}} = \sqrt{\frac{\alpha_T}{\beta}(T_{cr}(L) - T)}, \tag{13}$$

When compared Eqs.(7) and (9) with well-known expressions for bulk material, we obtained that $R_a(k=0)$ coincides with the transverse correlation length only under the condition $f = 1$, which is not the case for thin a-films. Moreover, the well-known expression for the critical thickness $L_{cr}$ of size-induced paraelectric phase transition cannot be obtained from the condition $R_a(0) \to \infty$. Instead it follows from the divergence of $\tilde{\chi}_{11}(0,z)$ in Eq.(9).

For considered a-film (see Appendix A for details) one should introduce z-dependent transverse correlation length in accordance with z-dependence of susceptibility $\tilde{\chi}_{11}(k,z)$ in Eq. (9) in the long wave-approximation ($\mathbf{k} = 0$) as [28]:

$$R_c^a(z) = \sqrt{g \cdot \tilde{\chi}_{11}(0,z)} = \sqrt{\frac{g \cdot f(z,L,R_a(0))}{\alpha(T) + 3\beta \overline{P}_1^2}}, \tag{14a}$$

$$\overline{R_c^a} \approx \sqrt{\frac{g}{3\beta \overline{P}_1^2 + \alpha_T(T - T_{cr}(L))}}. \tag{14b}$$

Thickness profile of correlation radius and generalized permittivity are shown in Figs.1. Thickness and temperature dependences of the permittivity and correlation radius averaged over the thickness of a-film are shown in Figs.2.

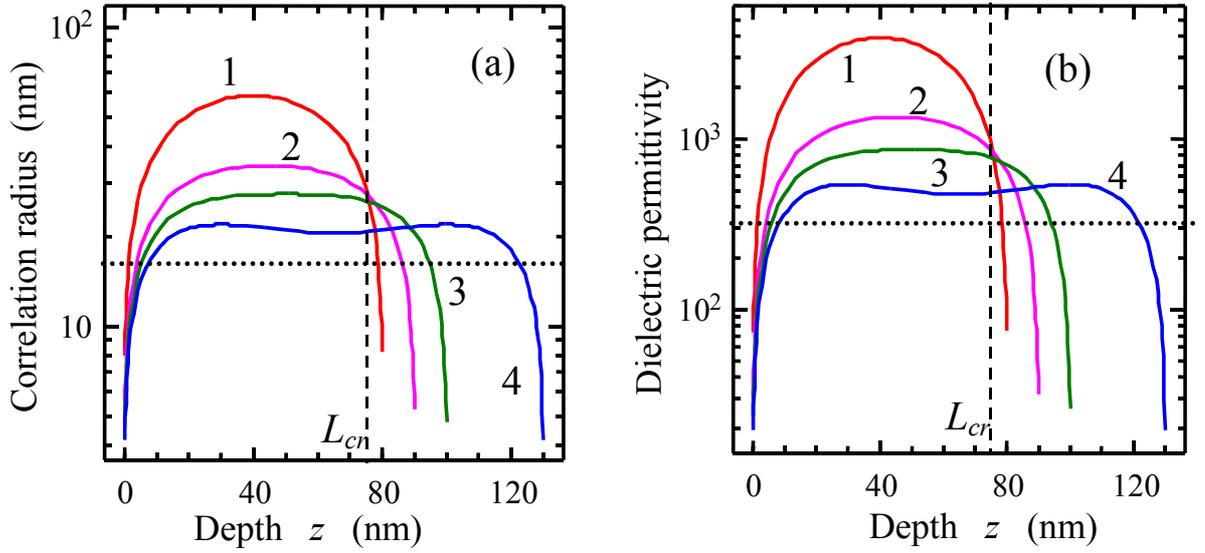

**Fig. 1**. Correlation radius (a) and dielectric permittivity (b) distribution in a-films of different thickness (curves 1-4 for $L$ = 80, 90, 100, 130 nm). Material parameters, $\alpha_T = 4.25 \cdot 10^5$ m/(F K), $T_C$ = 691 K, $\beta = 1.44 \cdot 10^8$ m$^5$/(C$^2$F), $g = 10^{-7}$ m$^3$/F, are typical for PbZr$_{0.4}$Ti$_{0.6}$O$_3$; $\lambda_{1,2}$ = 0.4 nm, Bulk values are marked by dotted lines.

One can see from Fig. 1, that correlation radius and permittivity profiles look like one another, although $R_c^a(z)$ has a little bit smeared maximum. The most important thing is the fact that correlation radius and permittivity increase with the film thickness decrease (but at $L>L_{cr}$), their maximal being much larger than in the bulk samples. Moreover, the condition $L \geq L_{cr} > R_c^a$ necessary for the phenomenological approach applicability is fulfilled for all 4 profiles shown in Fig. 1a.

It follows from Fig. 2 that the thickness dependence of average correlation radius and permittivity looks like their temperature dependences with divergences at critical thickness $L_{cr}(T)$ and critical temperatures $T_{cr}(L)$ respectively. So, the observations of correlation radius maxima will allow finding both these critical parameters.

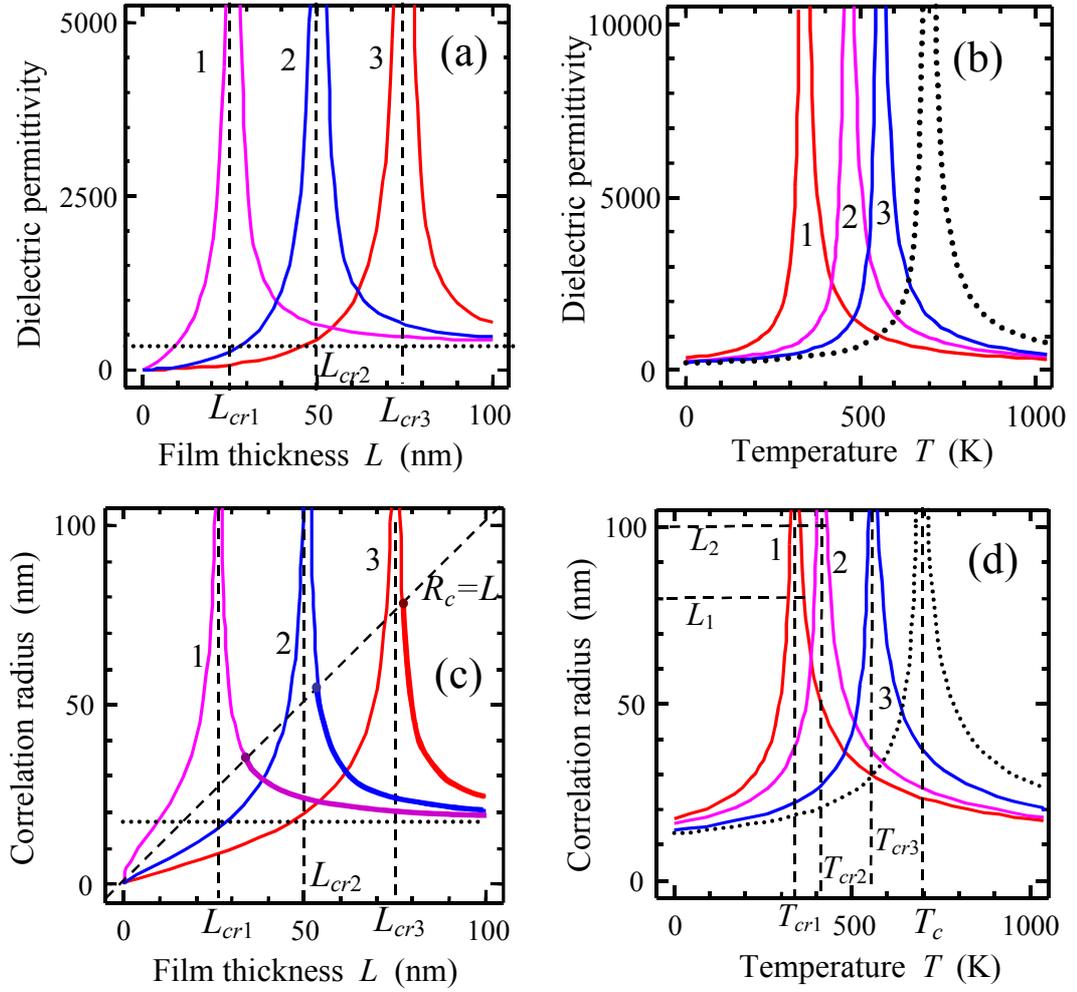

**Fig. 2**. Thickness (a, c) and temperature (b, d) dependences of average permittivity (a, b) and correlation radius (c, d) for the a-film. Material parameters are the same as in Fig. 1; (a, c) at room temperature for $\lambda_1 = 40$ nm, $\lambda_2 = 40$ nm (curve 1); $\lambda_1 = 0.4$ nm, $\lambda_2 = 40$ nm (curve 2); $\lambda_1 = 0.4$ nm, $\lambda_2 = 0.4$ nm (curve 3). (b, d) at $L = 80, 100, 130$ nm (curves 1, 2, 3 respectively) for $\lambda_1 = 0.4$ nm, $\lambda_2 = 0.4$ nm. Bulk system properties are shown by dotted curves.

On the other hand it is worth to underline that for all considered films with thickness $L > L_{cr}$ corresponding correlation radii are smaller than the film thickness $R_c^a < L$ below the bisector line in Fig. 2c (see bold curves), where the physically reasonable inequality $L \geq L_{cr} > R_c^a$ is valid and so the proposed criterion of LGD applicability is fulfilled for the considered films (see Figs. 1,2). The condition $R_c^a > L$ formally indicate the absolute stability of monodomain state in all the film, but it is clear from Fig. 2c, the inequality $R_c^a > L$ is valid only in the immediate vicinity of ferroelectric phase transition at $L = L_{cr}$, but allowing for high fluctuations in the critical region, this region to be excluded from phenomenological consideration similarly to what is known for bulk materials.

As for the temperature range for which LGD is valid, it is clear from Fig. 2d that it is almost all the range $T > T_{cr}(L)$ except the ultra-thin immediate vicinity of ferroelectric phase transition $T^*(L) > T > T_{cr}(L)$, where $R_c^a > L$ ($T^*$ is determined from the cross-section of the horizontal lines $R_c^a = L_{1,2,3}$ with corresponding curves 1,2,3).

## 4. Correlation radius and generalized susceptibility in c-films

*4.1. Solution for equilibrium spontaneous polarization distribution*

Using direct variational method, approximate analytical solution of Eq.(4) for c-films equilibrium polarization was derived as:

$$P_3(z) = P_V \cdot f(z, L, R_d), \quad P_V = \overline{P_3} + \frac{E_0 - \alpha \overline{P_3} - \beta \overline{P_3}^3 - (\eta/\varepsilon_0 \varepsilon_{33}^b) \overline{P_3}}{\alpha + 3\beta \overline{P_3}^2 + (1/\varepsilon_0 \varepsilon_{33}^b)}. \quad (15)$$

Where $f$ is given by Eq.(10), but the spatial scale $R_d$ is governed by:

$$R_d = \sqrt{\frac{\varepsilon_0 \varepsilon_{33}^b g}{\varepsilon_0 \varepsilon_{33}^b (\alpha + 3\beta \overline{P_3}^2) + 1}} \approx \sqrt{\varepsilon_0 \varepsilon_{33}^b g}. \quad (16)$$

Average polarization should be found self-consistently by the spatial averaging of Eq.(15), namely $\left(\varepsilon_0 \varepsilon_{33}^b \left(\alpha + 3\beta \overline{P_3}^2\right) + 1 - (1-\eta)\overline{f}\right) \overline{P_3} = \varepsilon_0 \varepsilon_{33}^b \left(2\beta \overline{P_3}^3 + E_0\right) \overline{f}$. For the case of no external field, $E_0=0$, we derived that in ferroelectric phase

$$\overline{P_3(L, R_d)} = \sqrt{\frac{-\varepsilon_0 \varepsilon_{33}^b \alpha - (1-\eta)\left(1 - \overline{f(L, R_d)}\right) - \eta}{\varepsilon_0 \varepsilon_{33}^b \beta \left(3 - 2\overline{f(L, R_d)}\right)}}. \quad (17)$$

Note that the critical thickness (as well as the critical temperature) of the size-induced phase transition should be found from zero denominator of Eq.(17). Namely, from the condition $\alpha(T) + \left(1 - \overline{f(L, R_d)}(1-\eta)\right)/\varepsilon_0 \varepsilon_{33}^b = 0$ we obtained:

$$T_{cr}(L) = T_C - \left(\frac{1-\eta}{\alpha_T \varepsilon_0 \varepsilon_{33}^b}\right) \frac{R_d^2 \left(2R_d (\cosh(L/R_d) - 1) + (\lambda_1 + \lambda_2) \sinh(L/R_d)\right)}{L(R_d(\lambda_1 + \lambda_2)\cosh(L/R_d) + (R_d^2 + \lambda_1\lambda_2)\sinh(L/R_d))} \quad (18)$$
$$\approx T_C - \left(\frac{1-\eta}{\alpha_T}\right) \frac{g(2R_d + \lambda_1 + \lambda_2)}{L(R_d(\lambda_1 + \lambda_2) + R_d^2 + \lambda_1\lambda_2)},$$

$$L_{cr}(T) \approx \frac{g(2R_d + \lambda_1 + \lambda_2)(1-\eta)}{R_d(\lambda_1 + \lambda_2) + R_d^2 + \lambda_1\lambda_2} \cdot \frac{1}{\alpha_T(T_C - T)}, \quad \text{at that} \quad T \leq T_C. \quad (19)$$

Note, that analytical expressions (17)-(19) with contribution of the screening parameter $\eta$ and different extrapolation length are derived for the first time.

*4.2. Generalized susceptibility and correlation radius in c-films*

Maxwell's equation lead to the integral relation between depolarization field and polarization fluctuations $\delta P_3(x,y)$ in Fourier **k**-representation as $\widetilde{E}_3^d\left[\delta\widetilde{P}_3\right]$ (see Eq.(B.1) in Appendix B). In k-domain we obtained that susceptibility $\widetilde{\chi}_{33}(k,z) = \dfrac{\delta\widetilde{P}_3(k,z)}{\delta\widetilde{E}_0(k)}$ satisfy the boundary problem

$$\begin{cases} \left(\alpha + 3\beta P_3^2(z) + gk^2\right)\widetilde{\chi}_{33} - g\dfrac{d^2}{dz^2}\widetilde{\chi}_{33} = 1 + \widetilde{E}_3^d\left[\widetilde{\chi}_{33}\right], \\ \left(\widetilde{\chi}_{33} - \lambda_1\dfrac{d}{dz}\widetilde{\chi}_{33}\right)\bigg|_{z=0} = 0, \quad \left(\widetilde{\chi}_{33} + \lambda_2\dfrac{d}{dz}\widetilde{\chi}_{33}\right)\bigg|_{z=L} = 0. \end{cases} \qquad (20)$$

The method of slow varying amplitudes leads to the approximate solution of inhomogeneous Eq.(20) (see Appendix A for more details). In particular, for small *k* values we obtained Pade approximation:

$$\chi_{33}(k,z) \approx \dfrac{f(z,L,R_d)}{gk^2 + \alpha(T) + 3\beta\overline{P}_3^2\left(3 - 2\overline{f(L,R_d)}\right) + \dfrac{1 - \overline{f(L,R_d)}(1-\eta)}{\varepsilon_0\varepsilon_{33}^b}}$$

$$= \begin{cases} \dfrac{f(z,L,R_d)}{gk^2 - 2\left(\alpha + \left(1 - \overline{f}(1-\eta)\right)/\varepsilon_0\varepsilon_{33}^b\right)}, & \text{FE phase} \\ \dfrac{f(z,L,R_d)}{gk^2 + \alpha + \left(1 - \overline{f}(1-\eta)\right)/\varepsilon_0\varepsilon_{33}^b}, & \text{PE phase} \end{cases} \qquad (21)$$

Here we neglect the smooth derivative $\partial\overline{f(L,R_d)}/\partial E_0$. So, using the same approach as for a-films, corresponding correlation radius in c-films could be defined as:

$$R_c^c(z) = \sqrt{g\cdot\widetilde{\chi}_{33}(0,z)} = \begin{cases} \sqrt{\dfrac{g\cdot f(z,L,R_d)}{-2\left(\alpha(T) + \left(1-(1-\eta)\overline{f(L,R_d)}\right)/\varepsilon_0\varepsilon_{33}^b\right)}}, & \text{FE phase} \\ \sqrt{\dfrac{g\cdot f(z,L,R_d)}{\alpha(T) + \left(1-(1-\eta)\overline{f(L,R_d)}\right)/\varepsilon_0\varepsilon_{33}^b}}, & \text{PE phase} \end{cases} \qquad (22)$$

Note, that in the limiting case $L \to \infty$ naturally $\overline{f(\infty,R_d)} \to 1$, $\eta \to 0$, so correlation length should be defined as $R_c^c(z) = \sqrt{\dfrac{g}{\alpha(T) + 3\beta\overline{P}_3^2}\left(1 - \dfrac{\exp(-z/R_d)}{1+\lambda_1/R_d}\right)}$, that exactly transforms into the bulk solution at $z \gg R_d$.

Thickness profile of correlation radius and generalized permittivity are shown in Figs.3. It is clear from (b) and (c) that the thickness of the region, where polarization gradient is essential is not less that 20 nm in the vicinity of both surfaces for the films with thickness larger than 80 nm, while correlation radius and permittivity are almost inhomogeneous in the whole 50-nm film. Note that Figs. 3c,d can be valid in the vicinity of another surface $z = L$ with substitution $L - z$ for $z$.

It is seen that similarly to a-films correlation radius and permittivity increase with c-film thickness decrease for thickness $L > L_{cr}$, their values being essentially higher than the ones in bulk samples (compare solid curves 4, 5, 6, 7). Situation is vise versa for dashed curves 1, 2, 3, which correspond to the films in paraelectric state at $L < L_{cr}$.

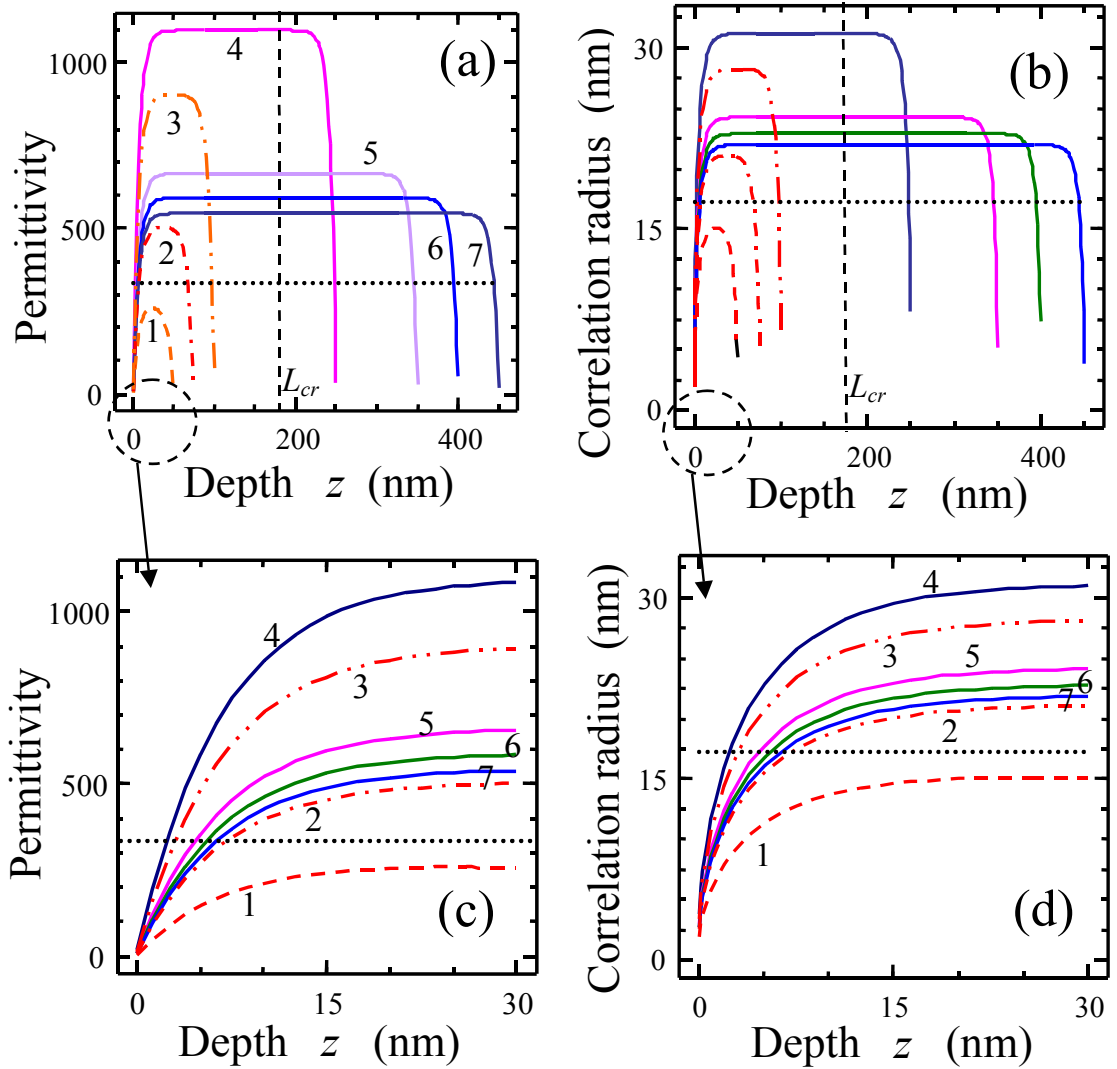

**Fig. 3**. Dielectric permittivity (a, c) and correlation radius (b, d) distributions in c-films of different thickness $L = 50, 75, 100, 250, 350, 400, 450$ nm (curves 1-7 respectively). (c, d) – zoomed region near z = 0. Parameter $\varepsilon_{33}^b = 50$, perfect electrodes with $R_{sc} = 0$. Material parameters, $\alpha_T = 4.25 \cdot 10^5$ m/(F K), $T_C = 691$ K, $\beta = 1.44 \cdot 10^8$ m$^5$/(C$^2$F), g = $10^{-7}$ m$^3$/F, are typical for PbZr$_{0.4}$Ti$_{0.6}$O$_3$; $\lambda_{1,2} = 0.1$ nm.

Thickness and temperature dependences of correlation radius averaged over the ***c-film*** depth are shown in Figs.4. In accordance with expression η valid for dielectric gaps, we introduced effective screening radius $R_{sc} = \varepsilon_{33}^b (H_1/\varepsilon_{g1} + H_2/\varepsilon_{g2})$, so that $\eta = \dfrac{R_{sc}}{R_{sc}+L}$.

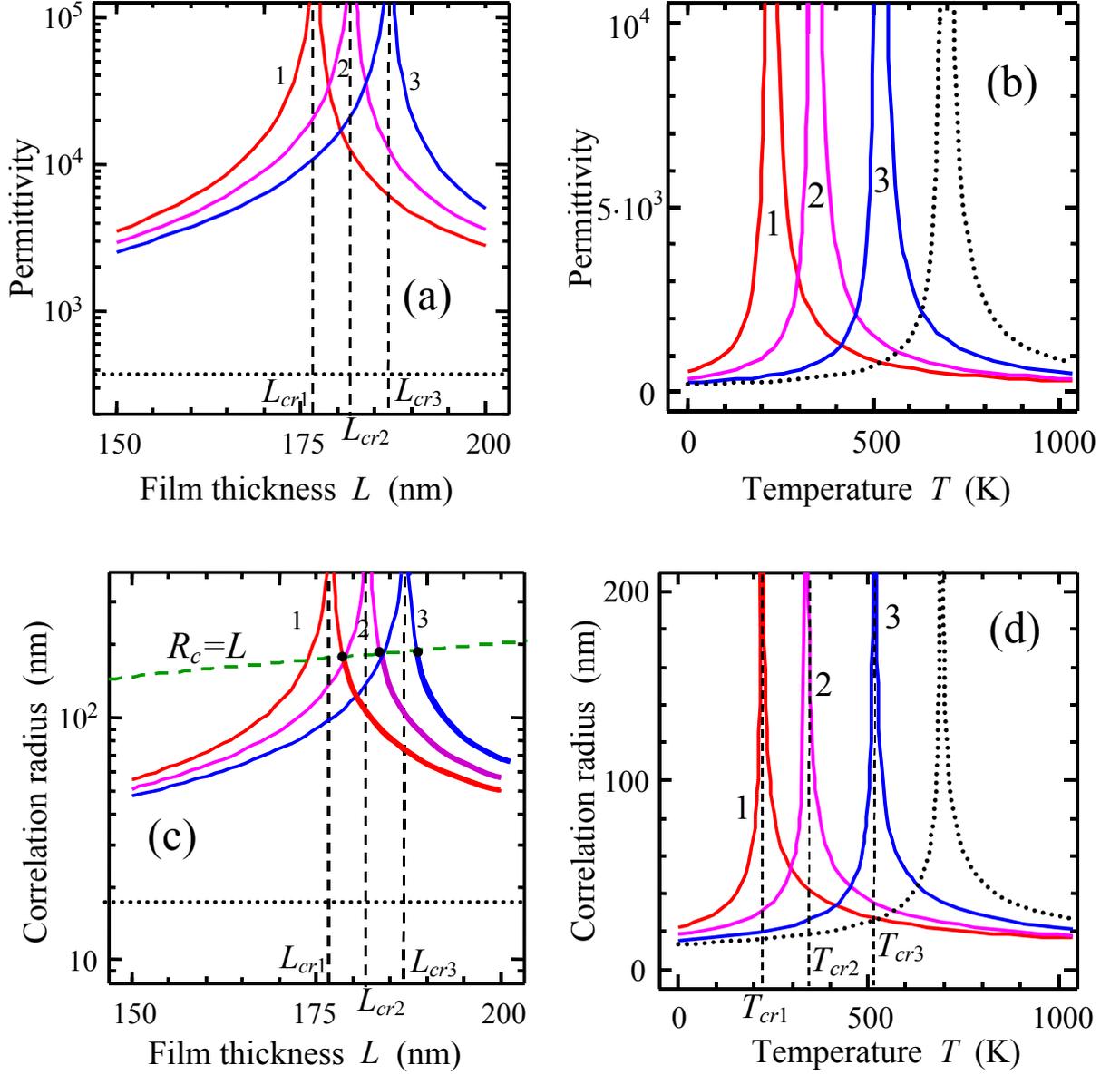

**Fig. 4**. Thickness (a, c) and temperature (b, d) dependences of average susceptibility (a, b) and correlation radius (c, d) for the c-film. Material parameters are the same as in Fig. 1; (a, c) at room temperature for $R_{sc}$ = 0, 0.4, 0.8 nm (curves 1, 2, 3 respectively). (b, d) at $L$ = 150, 200, 400 nm (curves 1, 2, 3 respectively). $\lambda_1$ = 0.1 nm, $\lambda_2$ = 0.2 nm, $\varepsilon_{33}^b$ = 50. Bulk system properties are shown by dotted curves.

The divergence of the average correlation radii at critical temperatures and film thicknesses is clearly seen from Figs. 4, at that the critical temperature depends on the film thickness (as well as the critical thickness depends on the temperature). For considered case of positive extrapolation lengths the critical temperature decreases with the film thickness decrease.

One can see from Fig. 4c that in the thickness region of ferroelectric phase ($L > L_{cr}$) the correlation radius of bulk sample (dotted line) is much smaller than in the films. Similarly to the situation discussed in the section 4.1 for a-films, for all considered c-films with thickness $L > L_{cr}$ corresponding correlation radii are smaller than the film thickness $R_c^c < L$ below the dashed line $R_c^c = L$ in Fig. 4c (see bold curves), where the physically reasonable inequality $L \geq L_{cr} > R_c^c$ is valid and so the proposed criterion of LGD applicability is fulfilled for the considered films in ferroelectric phase.

**Discussion**

For the second order ferroelectrics the correlation radius and generalized susceptibility diverge at the film critical thickness and temperature as anticipated. It is worth to underline that in the most part of film depth (except the small sub-surface region in c-films) correlation radius of confined system are higher than the bulk correlation radius (see Figs.1 and 3). The depth profiles of correlation radius and generalized susceptibility in c-films with strong depolarization field have much broader region or plato-like behavior in comparison with the ones in a-films without depolarization field. For thick a-films maxima near the film surfaces appeared allowing for strong polarization gradient in the region. The maxima are suppressed by inevitable depolarization effects in c-films (compare Figs.1 and 3). However the depth of "shell" regions near both surfaces is well above the lattice constant even in thick c-film (see Figs.3c,d). Minimal depolarization field related with intrinsic polarization gradient conserves in c-film even for the perfect electrodes with $\eta = 0$, contrary to bulk samples. The effect lead to the size-induced phase transition itself (see e.g. Figs.3a,b). The increase of $\eta$ leads to the critical thickness increase (see e.g. Figs.3c,d).

The critical thickness in a-films is essentially smaller than the one for c-films allowing for the absence of depolarization field in a-films. However for both thin a- and c-films in ferroelectric phase ($L > L_{cr}$) correlation radius are typically much smaller than the film thickness (see Figs.1,3), it diverges only in the immediate vicinity of the critical thickness/critical temperature. In all realistic cases of ferroelectric phase existence in films of thickness $L$ we obtained that $L \geq L_{cr} > R_c$ is fulfilled, proving that the applicability of phenomenological approach is out of doubts.

Obtained results prove that oversimplifications of the phenomenological approach (such as to neglect the polarization gradient and/or unjustified assumption about negligibly small surface energy contribution) lead to invalid conclusions about the contributions of different physical mechanisms into the size effect of thin ferroelectric films.

**Acknowledgements**

Research sponsored by Ministry of Science and Education of Ukrainian and National Science Foundation (Materials World Network, DMR-0908718).

**Appendix A. a-films**

Green function can be rewritten in two identical representation as

$$G(R,z,z') = \frac{R}{g} \cdot \frac{(\lambda_1 \cosh(z/R) + R\sinh(z/R))(\lambda_2 \cosh((L-z')/R) + R\sinh((L-z')/R))}{R(\lambda_1 + \lambda_2)\cosh(L/R) + (R^2 + \lambda_1\lambda_2)\sinh(L/R)}$$
$$\equiv \sum_n \frac{(q_n\lambda_1 \cos(q_n z) + \sin(q_n z))(q_n\lambda_1 \cos(q_n z') + \sin(q_n z'))}{g(R^{-2} + q_n^2)N(q_n)} \quad (A.1)$$

The summation is performed on the values of eigen values $q_n$, which in turn should be found from the following condition $q_n(\lambda_1 + \lambda_2)\cos(q_n L) + (1 - q_n^2\lambda_1\lambda_2)\sin(q_n L) = 0$. Under this condition the eigen functions $q_n\lambda_1\cos(q_n z) + \sin(q_n z)$ satisfies the boundary conditions (5). $N(q_n)$ is the eigen-functions norm $N(q) = [2q(L + \lambda_1 + Lq^2\lambda_1^2 - \lambda_1\cos(2Lq)) - (1 - q^2\lambda_1^2)\sin(2qL)](4q)^{-1}$.

Generalized susceptibility defined from Eqs.(6) as variational derivative leads to the expression

$$\tilde{\chi}_{11}(k,z) = \frac{\partial \delta \tilde{P}_1(\mathbf{k},z)}{\partial \delta \tilde{E}_1(\mathbf{k})} = \int_0^z dz' G(R_a(k),z,z') + \int_z^L dz' G(R_a(k),z',z) = \frac{R_a^2(k)}{g} f(z,L,R_a(k))$$
$$\equiv \sum_n \frac{(1 + q_n\lambda_1\sin(q_n L) - \cos(q_n L))(q_n\lambda_1\cos(q_n z) + \sin(q_n z))}{g(R_a^{-2}(q_n) + k^2) \cdot q_n \cdot N(q_n)} \quad (A.2)$$

The expression for the critical thickness $L_{cr}(T)$ of size-induced paraelectric phase transition can be obtained from the divergence of $\tilde{\chi}_{11}(0,z)$ at $\overline{P}_1^2 \to 0$, namely:

$$L_{cr}(T) = \sqrt{\frac{g}{\alpha}}\operatorname{arctanh}\left(-\frac{\sqrt{g/\alpha}(\lambda_1+\lambda_2)}{\lambda_1\lambda_2+g/\alpha}\right) \equiv \sqrt{\frac{-g}{\alpha(T)}}\left(\arctan\left(\sqrt{-\frac{g}{\alpha}}\frac{1}{\lambda_1}\right) + \arctan\left(\sqrt{-\frac{g}{\alpha}}\frac{1}{\lambda_2}\right)\right) \quad (A.3)$$

Note, that $R_a(q_1)$ diverges at $L = L_{cr}$, since $q_1^2(L_{cr}) = -\alpha(T)/g$. The fact leading to the divergence of the series in Eq.(A.2) under the condition $\mathbf{k} = 0$. Also we obtained that

$$R_a(q_1) = \sqrt{\frac{g}{\alpha(T) + 3\beta\overline{P}_1^2 + gq_1^2}} \approx \sqrt{\frac{g}{3\beta\overline{P}_1^2 + \alpha_T(T - T_{cr}(L))}}. \quad (A.3)$$

So, since each term of the series in Eq.(A.2), proportional to $(R_a^{-2}(q_n) + k^2)^{-1} \sim (\alpha(T) + 3\beta\overline{P}_1^2 + g(k^2 + q_n^2))^{-1}$ has the same functional form as the 3D-Fourier image of the Ornshtein-Zernike correlation function in bulk material, expression (A.3) may be considered as approximation for correlation radius. Rigorously speaking for considered a-film one should consider either the infinite series of $\{R_a(q_n)\}$, or to introduce z-dependent transverse correlation length in accordance with z-dependence of susceptibility $\tilde{\chi}_{11}(k,z)$ in Eq. (9) in the long wave-approximation (**k** = 0).

**Appendix B. c-films**

Maxwell's equation lead to the integral relation between depolarization field and polarization fluctuations $\delta P_3(x,y)$ in Fourier **k**-representation. Depolarization field in c-film was derived in Ref.[38] as

$$\tilde{E}_3^d(\mathbf{k},z) = -\frac{k}{\gamma}\int_0^L d\xi \frac{\cosh(k(L-\xi)/\gamma)\delta\tilde{P}_3(\mathbf{k},\xi)\sinh(kH)\cosh(k(L-z)/\gamma)}{\varepsilon_0\left(\varepsilon_{33}^b\cosh(kL/\gamma)\sinh(kH)+\gamma\varepsilon_g\cosh(kH)\sinh(kL/\gamma)\right)\sinh(kL/\gamma)}$$

$$-\frac{\delta\tilde{P}_3(\mathbf{k},z)}{\varepsilon_0\varepsilon_{33}^b} + \frac{k}{\gamma}\left(\begin{array}{l}\int_0^z d\xi \dfrac{\cosh(k\xi/\gamma)\cosh(k(L-z)/\gamma)}{\varepsilon_0\varepsilon_{33}^b \cdot \sinh(kL/\gamma)}\delta\tilde{P}_3(\mathbf{k},\xi) \\ +\int_z^L d\xi \dfrac{\cosh(kz/\gamma)\cosh(k(L-\xi)/\gamma)}{\varepsilon_0\varepsilon_{33}^b \cdot \sinh(kL/\gamma)}\delta\tilde{P}_3(\mathbf{k},\xi)\end{array}\right) \quad \text{(B.1)}$$

Here $\gamma = \sqrt{\varepsilon_{33}^b/\varepsilon_{11}}$, vector $\mathbf{k} = \{k_1,k_2\}$ and its absolute value is $k = \sqrt{k_1^2+k_2^2}$.

Allowing for Maxwell's equations $\left(\dfrac{d^2}{dz^2}-\dfrac{k^2}{\gamma^2}\right)\delta\tilde{E}_3(\mathbf{k},z) = -\dfrac{d^2}{dz^2}\dfrac{\delta\tilde{P}_3(\mathbf{k},z)}{\varepsilon_0\varepsilon_{33}^b}$, where $\delta\tilde{E}_3(\mathbf{k},z) = \delta\tilde{E}_0(\mathbf{k},z) + \tilde{E}_3^d(\mathbf{k},z)$. Thus, applying the Chensky-Tarasenko operator $\dfrac{d^2}{dz^2} - \dfrac{k^2}{\gamma^2}$ to Eq.(20), we obtained 4th order differential equation with z-dependent coefficients

$$\left(\frac{d^2}{dz^2}-\frac{k^2}{\gamma^2}\right)\left(\left(\alpha+3\beta P_3^2(z)+gk^2\right)\tilde{\chi}_{33}-g\frac{d^2}{dz^2}\tilde{\chi}_{33}\right) = -\frac{d^2}{dz^2}\frac{\tilde{\chi}_{33}}{\varepsilon_0\varepsilon_{33}^b}. \quad \text{(B.2)}$$

The method of slow varying amplitudes leads to the approximate solution of inhomogeneous Eq.(B.2).

In particular case $k\to 0$ (long wave-approximation) *exact solution* for averaged dielectric susceptibility was derived self-consistently from Eq.(15) rewritten for the susceptibility $\overline{\chi}_{33} = d\overline{P}_3/dE_0$ as $\left(\alpha + \dfrac{1-(1-\eta)\overline{f}}{\varepsilon_0\varepsilon_{33}^b}\right)\overline{\chi}_{33} + 9\beta\overline{P}_3^2\cdot\overline{\chi}_{33} = \left(6\beta\overline{P}_3^2\cdot\overline{\chi}_{33}+1\right)\overline{f}$. The method of slow varying amplitudes leads to the approximate solution of inhomogeneous Eq.(B.2) in the form

$$\tilde{\chi}_{33}(k,z) \approx p_0 + B_1\exp(s_1(k)z) + C_1\exp(-s_1(k)z) + B_2\exp(s_2(k)z) + C_2\exp(-s_2(k)z) \quad \text{(B.3)}$$

The expression for $p_0$ should be found self-consistently from Eq.(20) allowing for depolarization field Eq.(B.1).

After the substitution $P_3^2(z)\to\overline{P}_3^2$, values $s_1(k)$ and $s_2(k)$ one can find as the roots of the characteristic equation, namely

$$\left(s^2 - \frac{k^2}{\gamma^2}\right)\left(\alpha + 3\beta\overline{P}_3^2 + g(k^2 - s^2)\right) + \frac{s^2}{\varepsilon_0 \varepsilon_{33}^b} = 0,$$

$$s_{1,2}^2 = \frac{1}{2}\left(\frac{(\alpha + 3\beta\overline{P}_3^2 + gk^2)}{g} + \frac{k^2}{\gamma^2} + \frac{1}{\varepsilon_0 \varepsilon_{33}^b g}\right) \pm \quad \text{(B.4)}$$

$$\pm \frac{1}{2}\sqrt{\left(\frac{(\alpha + 3\beta\overline{P}_3^2 + gk^2)}{g} + \frac{k^2}{\gamma^2} + \frac{1}{\varepsilon_0 \varepsilon_{33}^b g}\right)^2 - \frac{4k^2}{\gamma^2 g}(\alpha + 3\beta\overline{P}_3^2 + gk^2)}$$

At $k = 0$ we obtained that $s_2 = 0$, $s_1 \approx R_d^{-1} \approx \sqrt{1/(\varepsilon_0 \varepsilon_{33}^b g)}$.

At $k \neq 0$ the condition of identically zero coefficients before $\cosh(k(L-z)/\gamma)$ and $\cosh(kz/\gamma)$ lead to

$$\sum_{i=1}^{2} \frac{s_i(B_i - C_i)}{(k^2 - \gamma^2 s_i^2)} = \sum_{i=1}^{2} \frac{s_i(B_i \exp(s_i L) + C_i \exp(-s_i L))}{(k^2 - \gamma^2 s_i^2)} = 0 \quad \text{(B.5)}$$

Expressing constants $B_i$ via $C_i$ we obtained that

$$B_1 = \frac{C_1 s_1(k^2 - \gamma^2 s_2^2)(\exp(s_2 L) + \exp(-s_1 L)) + C_2 s_2(k^2 - \gamma^2 s_1^2)(\exp(-s_2 L) + \exp(s_2 L))}{s_1(k^2 - \gamma^2 s_2^2)(\exp(s_2 L) - \exp(s_1 L))}, \quad \text{(B.6a)}$$

$$B_2 = \frac{C_2 s_2(k^2 - \gamma^2 s_1^2)(\exp(s_1 L) + \exp(-s_2 L)) + C_1 s_1(k^2 - \gamma^2 s_2^2)(\exp(-s_1 L) + \exp(s_1 L))}{s_2(k^2 - \gamma^2 s_1^2)(\exp(s_1 L) - \exp(s_2 L))}. \quad \text{(B.6b)}$$

Two constants $C_{1,2}$ should be found from the boundary conditions to Eq.(20).

In particular case $L \to \infty$ that one should put $B_1 = B_2 = 0$ in Eq.(B.3) that is true under the condition $C_1 s_1(k^2 - \gamma^2 s_2^2) + C_2 s_2(k^2 - \gamma^2 s_1^2) = 0$. So $C_2 = -C_1 \frac{s_1(k^2 - \gamma^2 s_2^2)}{s_2(k^2 - \gamma^2 s_1^2)}$. Finally the remained boundary condition at z=0 gives equation for the constant $C_1$, namely: $p_0(k) + (1 + \lambda_1 s_1)\left(1 - \frac{(k^2 - \gamma^2 s_2^2)}{(k^2 - \gamma^2 s_1^2)}\right) C_1 = 0$. Naturally that $\overline{f(\infty, R_d)} \to 1$, $\eta \to 0$. So the solution has the form

$$\chi_{33}(k,z) = \frac{1}{\alpha + 3\beta\overline{P}_3^2 + gk^2}\left(1 - \frac{\exp(-s_1 z) - \frac{s_1(k^2 - \gamma^2 s_2^2)}{s_2(k^2 - \gamma^2 s_1^2)}\exp(-s_2 z)}{(1 + \lambda_1 s_1)\left(1 - \frac{(k^2 - \gamma^2 s_2^2)}{(k^2 - \gamma^2 s_1^2)}\right)}\right) \quad \text{(B.7)}$$

Corresponding generalized susceptibility is $\widetilde{\chi}_{33}(k,z) = \left.\frac{\delta \widetilde{P}_3}{\delta \widetilde{E}_0}\right|_{\delta \widetilde{E}_0 \to 0}$.

[37] Depolarization field should be chosen zero as $E_S^d(\delta P_1(y,z)) = E_S^d(\delta P_2(x,z)) = 0$.